\documentclass[paper]{myJHEP3}
\usepackage{axodraw}
\usepackage[T1]{fontenc} % if needed
\title{\boldmath On the predictivity of the non-renormalizable quantum field theories}

\author{R. Pittau \\
Departamento de F\'{\i}sica Te\'orica y del Cosmos and CAFPE,\\
Campus Fuentenueva s. n., Universidad de Granada 
E-18071 Granada, Spain \\
E-mail: \email{pittau@ugr.es}}

\abstract{Following a Four Dimensional Renormalization approach to ultraviolet divergences (FDR), we extend the concept of predictivity to non-renormalizable quantum field theories at arbitrarily large perturbative orders. The idea of {\em topological renormalization} is introduced, which keeps a finite value for the parameters of the theory by trading the usual order-by-order renormalization procedure for an order-by-order redefinition of the perturbative vacuum.  One additional measurement is then sufficient to systematically compute quantum corrections at any loop order, with no need of absorbing 
ultraviolet infinities in the Lagrangian.}

\preprint{}

\begin{document} 

\newcommand{\mur}{\mu_{\scriptscriptstyle R}}
\newcommand{\bqa}{\begin{eqnarray}}
\newcommand{\eqa}{\end{eqnarray}}
\newcommand{\D}{\,{D}}
\newcommand{\Dbar}{\bar D}
\newcommand{\dbar}{\bar D}
\newcommand{\qbar}{\bar q}
\newcommand{\nl}{\nonumber \\}
\maketitle
\flushbottom
\section{Introduction}
\label{sec:intro}
Renormalizability~\cite{Bogoliubov:1957gp,Hepp:1966eg,Zimmermann:1969jj} provides a very powerful guideline in high-energy particle physics.  For instance, the existence of intermediate vector bosons in the Electroweak Standard Model (SM) can be directly inferred from the non-renormalizability of the low-energy four-fermion interaction. A second example is the pattern of the SM fermions, which is very strongly constrained by the need of canceling the gauged axial anomaly.
A renormalizable Quantum Field Theory (QFT) is {\em complete}. After fixing its free parameters, any observable can be predicted at any perturbative order and scale. The ultraviolet (UV) infinities occurring in the intermediate steps of the calculation~\cite{Dyson:1949ha,Dyson:1949bp,Salam:1951sm,Salam:1951sj}, being universal, are re-absorbed in the parameters of the theory and any UV cutoff disappears, leaving a dependence on the renormalization scale at perturbative orders higher than the computed one. 

On the contrary, in non-renormalizable theories new infinities are generated, order by order in the loop expansion, which cannot be re-absorbed into the original Lagrangian. Such theories are usually interpreted 
as {\em effective} ones~\cite{Weinberg:1978kz}, and the traditional renormalization program works untouched only if the original Lagrangian is modified, at any given order, by adding a new set of interactions, where the new generated infinities can be accommodated. The ignorance of the UV completion coming from the {\em true} theory is parametrized by a change in the original Lagrangian. However, the new interactions need to be fixed by experiment, which leads, in principle, to a lack of predictivity of the theory at arbitrarily large perturbative orders. 

In this work, we use the FDR approach introduced in~\cite{Pittau:2012zd} to discuss a possible way out to this problem. The basic idea is as follows: the type and number of UV infinities occurring at any loop order is reduced and kept under control once they are classified in terms of their {\em topology} at the {\em integrand} level\footnote{We work in momentum space.}, rather than according to their occurrence in the interactions. Thus, they can be subtracted from the physical spectrum without the need of re-absorbing them in the Lagrangian's parameters, which are left untouched. 
This subtraction can be interpreted as an order-by-order redefinition of the perturbative vacuum, dubbed  {\em topological renormalization}, in which non-physical configurations are discarded. 

In a non-renormalizable QFT this separation between physical and non physical degrees of freedom may cause the appearance of an arbitrary logarithmic parameter {\em at the same perturbative level} one is calculating, that can be interpreted as the scale at which the UV divergent configurations are subtracted. 
To pull such a dependence back to higher perturbative orders one additional measurement is necessary, which links it to the parameters of the theory. In this way, the (unknown) high-energy UV behavior can be completely decoupled from the low energy regime. The advantages are clear. Even though non-renormalizabe theories remain --\,technically\,-- effective, the original symmetries are kept (since the Lagrangian is never touched), and finiteness and predictivity are both restored. 

Following this interpretation, interactions must be introduced in a non-renormalizable bare Lagrangian based solely on physical motivation, and not on the need of absorbing UV divergences. Thus, any effective Lagrangian~\cite{Pich:1998xt} containing only a subset of all possible higher dimensional operators represents a legitimate QFT, whose validity has to be judged a-posteriori by comparing its predictions with data.

The outline of the paper is as follows. Section~\ref{sec:fdr} reviews the aspects of FDR which are relevant to the present discussion. In Sec.~\ref{sec:clas} a classification of the UV infinities occurring in scalar and tensor loop integrals is presented and the {topological renormalization} is discussed. Finally, Sec.~\ref{sec:renandnonren} analyzes the use of FDR in non-renormalizable theories compared to the renormalizable case.

\section{The FDR integration}
\label{sec:fdr}
In FDR~\cite{Pittau:2012zd,Pittau:2013qla,Donati:2013iya,Donati:2013voa} the UV infinities are subtracted at the {\em integrand} level by judiciously splitting the original integrand $J(q_1,\ldots, q_\ell)$ of an $\ell$-loop function\footnote{$q_1,\ldots, q_\ell$ are integration momenta and $J(q_1,\ldots, q_\ell)$ can be a tensor.} in two parts, 
$J_{\rm INF}(q_1,\ldots, q_\ell)$ and $J_{{\rm F},\ell}(q_1,\ldots, q_\ell)$:
\bqa
\label{eq:fdrexpansion}
J(q_1,\ldots, q_\ell)= J_{\rm INF}(q_1,\ldots, q_\ell)+J_{{\rm F},\ell}(q_1,\ldots, q_\ell).
\eqa
The former piece collects integrands which would produce UV divergences upon integration ({\em divergent  integrands}) and is discarded, while the latter is kept and generates the physical finite contribution. To regulate the spurious infrared (IR) divergences caused by this \mbox{break-up}\footnote{The sum $J_{\rm INF}+J_{\rm F,\ell}$ is free of spurious IR poles while $J_{\rm F,\ell}$ alone is not.}, the $+i 0$ propagator prescription is made explicit by identifying it with a vanishing mass $-\mu^2$ and taking the limit $\mu \to 0$ outside integration. The rationale for this separation is that the divergent integrands in $J_{\rm INF}(q_1,\ldots, q_\ell)$ are allowed to depend on $\mu$, {\em but not on physical scales}, so that physics is entirely contained in $J_{{\rm F},\ell}(q_1,\ldots, q_\ell)$. As a two-loop example consider the rank-2 irreducible tensor
\bqa
J^{\alpha \beta}(q_1,q_2) = \frac{q^\alpha_1 q^\beta_1}{\dbar^3_1 \dbar_2 \dbar_{12}},
\eqa
with
\bqa
\label{eq:2lden}
\bar D_1   = \bar{q}_1^2-m_1^2,~~
\bar D_2   = \bar{q}_2^2-m_2^2,~~
\bar D_{12} = \bar{q}_{12}^2-m_{12}^2\,,~~
q_{12}= q_1+q_2\,,~~\bar q^2_j = q^2_j-\mu^2. 
\eqa
The desired splitting can be obtained by means of a repeated use of the identities
\bqa
\label{eq:ids}
\frac{1}{\dbar_j}  = \frac{1}{\qbar^2_j} +\frac{m^2_j}{\qbar^2_j \dbar_j},~~~~
\frac{1}{\qbar^2_{12}}= \frac{1}{\qbar^2_2}-\frac{q^2_1+2 (q_1 \cdot q_2)}{\qbar^2_2 \qbar^2_{12} },~~~~
\frac{1}{\qbar^2_{2}}= \frac{1}{\qbar^2_1}-\frac{q^2_{12}-2 (q_1 \cdot q_{12})}{\qbar^2_1 \qbar^2_2 },
\eqa
and reads
\bqa
\label{eq:appa1}
J^{\alpha \beta}(q_1,q_2) &=&
q^\alpha_1 q^\beta_1 
\left\{
\left[\frac{1}{\qbar^6_1 \qbar^2_2 \qbar^2_{12}}  \right]
  +\left(
        \frac{1}{\bar D^3_1}
       -\frac{1}{\bar{q}_1^6}
    \right)
\left(
 \left[\frac{1}{\qbar_2^4}\right]
 -\frac{q_1^2+2(q_1 \cdot q_2)}{\bar{q}_2^4 \bar{q}_{12}^2} 
\right)
\right. \nl
 &&+\left.\frac{1}{\bar D_{1}^3 \bar{q}_2^2\bar D_{12} }
 \left(
 \frac{m_2^2}{\bar D_{2}}+\frac{m_{12}^2}{\bar{q}_{12}^2}
 \right) 
\right \},
\eqa
where divergent integrands are written between square brackets.
Therefore
\bqa
\label{eq:ex2l}
J^{\alpha \beta}_{\rm INF}  (q_1,q_2) &=& 
q^\alpha_1 q^\beta_1 
\left\{
\left[\frac{1}{\qbar^6_1 \qbar^2_2 \qbar^2_{12}}  \right]
  +\left(
        \frac{1}{\bar D^3_1}
       -\frac{1}{\bar{q}_1^6}
    \right)
 \left[\frac{1}{\qbar_2^4}\right]
\right \}\,~~~{\rm and} \nl
J^{\alpha \beta}_{{\rm F},2}(q_1,q_2) &=& 
q^\alpha_1 q^\beta_1 
\left\{
 \frac{1}{\bar D_{1}^3 \bar{q}_2^2\bar D_{12} }
  \left(
  \frac{m_2^2}{\bar D_{2}}+\frac{m_{12}^2}{\bar{q}_{12}^2}
  \right) 
  -\left(
        \frac{1}{\bar D^3_1}
       -\frac{1}{\bar{q}_1^6}
    \right)
 \frac{q_1^2+2(q_1 \cdot q_2)}{\bar{q}_2^4 \bar{q}_{12}^2} 
\right\}.
\eqa 

The FDR integral over the original integrand $J(q_1,\ldots, q_\ell)$ is {\em defined}, through Eq.~(\ref{eq:fdrexpansion}), as\footnote{FDR integration is denoted by the symbol $[d^4q_i]$.}
\bqa
\label{eq:fdrdef}
\int [d^4q_1] \ldots [d^4q_\ell]\,  J(q_1,\ldots, q_\ell) \equiv \lim_{\mu \to 0}
\int d^4q_1 \ldots d^4q_\ell\, J_{{\rm F},\ell}(q_1,\ldots q_\ell),
\eqa
and the expansion needed to extract $J_{{\rm F},\ell}(q_1,\ldots q_\ell)$ is called the {\em FDR defining expansion} of $J(q_1,\ldots, q_\ell)$.
For instance, from Eq.~(\ref{eq:ex2l})
\bqa
\label{eq:exexex}
\int [d^4q_1] [d^4q_2] \frac{q^\alpha_1 q^\beta_1}{\dbar^3_1 \dbar_2 \dbar_{12}}
= \lim_{\mu \to 0} \int d^4q_1 d^4q_2\,J^{\alpha \beta}_{{\rm F},2}(q_1,q_2).
\eqa
Notice that, as anticipated, the IR behavior 
$J^{\alpha \beta}_{{\rm F},2}(q_1,q_2)~~{\sim}^{\!\!\!\!\!\!\!\!\!^{q_i \to\, 0}}\, \frac{1}{\qbar^4_i}$ is regulated by $\mu^2$. In convergent integrals $J_{\rm INF}(q_1,\ldots q_\ell)= 0$, thus FDR integration and normal integration coincide. Conversely, polynomials in the integration variables represent a limiting case of Eq.~(\ref{eq:fdrexpansion}), in which $J_{{\rm F},\ell}(q_1,\ldots, q_\ell)= 0$. As a consequence
\bqa
\int [d^4q] \left(\qbar^2\right)^\alpha = 0\,,
\eqa
for any integer $\alpha \geq 0$.

The FDR integration in Eq.~(\ref{eq:fdrdef}) encodes the UV subtraction {\em directly} into its definition and satisfies, at the same time, the two mathematical properties required for (regulated) divergent integrals to maintain the symmetries of the QFT at hand (including gauge invariance\footnote{We rely here on the existence of graphical proofs of the Ward-Slavnov-Taylor identities~\cite{Sterman:1994ce}, in which the correct relations among Green's functions are demonstrated diagrammatically --\,at any loop order\,-- 
by means of algebraic manipulations of the 
{\em integrands} of the loop functions.}), namely:
\begin{itemize}
\item[i)]invariance under shift of any integration variable~\cite{Collins:1984xc}
\bqa
\label{eq:shift}
\int [d^4q_1] \ldots [d^4q_\ell]\,  J(q_1,\ldots, q_\ell) &=& 
\int [d^4q_1] \ldots [d^4q_\ell]\,  J(q_1+p_1,\ldots, q_\ell+p_\ell) \nl
  \forall\, p_i,~{\rm with}~ i &=& 1,\ldots,\ell;  
\eqa
\item[ii)] preservation of the cancellations between numerators and 
denominators~\cite{veltman74}
\bqa
\label{eq:canc}
\int [d^4q_1] \ldots [d^4q_\ell]\, \frac{\qbar^2_i-m^2_i}{ (\qbar^2_i-m^2_i)^m \ldots}
= 
\int [d^4q_1] \ldots [d^4q_\ell]\, \frac{1}{ (\qbar^2_i-m^2_i)^{m-1}\ldots} .
\eqa
\end{itemize}
The first property can be proven\footnote{See Appendix A of~\cite{Donati:2013voa} for more details.} by rewriting FDR integrals as a finite difference of UV divergent integrals regulated in Dimensional Regularization (DR)\footnote{Here and in the following $n= 4+ \epsilon$ and $\mur$ is the arbitrary scale of DR.}
\bqa
\label{eq:diff}
\int [d^4q_1] \ldots [d^4q_\ell]\,  J(q_1,\ldots, q_\ell) &=& 
\lim_{\mu \to 0} \mur^{-\ell \epsilon} \left( 
  \int d^nq_1 \ldots d^nq_\ell \, J(q_1,\ldots, q_\ell) \right. \nl
&& -\left. \int d^nq_1 \ldots d^nq_\ell \, J_{\rm INF}(q_1,\ldots, q_\ell)
       \right),
\eqa 
and it follows from the fact that the r.h.s. of Eq.~(\ref{eq:diff}) is shift invariant. The second property holds if a replacement\footnote{Only one kind of $\mu^2$ exists. The index $_i$ in $\mu^2|_i$ only denotes that the denominator expansion in front of $\mu^2$ should be {\em the same one used for $q^2_i$} when it appears in the numerator of an integral, as in Eq.~(\ref{eq:mu}).}
\bqa
\label{eq:glopresc}
q^2_i \to \qbar^2_i= q^2_i-\mu^2|_i
\eqa
is performed for any $q^2_i$ generated by Feynman rules.\footnote{The replacement in Eq.~(\ref{eq:glopresc}) is called {\em global prescription}.} 
For instance, from the defining expansions of the two integrands
\bqa
\label{2loopFDR}
\int [d^4q_1] [d^4q_2] 
  \frac{q^2_1 -\mu^2|_1-m_1^2}{\bar D_1^3\bar D_2\bar D_{12}} =
\int [d^4q_1] [d^4q_2] 
  \frac{1}{\bar D_1^2\bar D_2\bar D_{12}},
\eqa
as long as the integral containing $\mu^2|_1$ is {\em defined} as
\bqa
\label{eq:mu}
&&\int [d^4q_1] [d^4q_2] 
 \frac{\mu^2|_1}{\bar D_1^3\bar D_2\bar D_{12}} \nl 
&&~~~ \equiv \lim_{\mu \to 0}\, \mu^2 \int d^4q_1 d^4q_2 
\left\{
 \frac{1}{\bar D_{1}^3 \bar{q}_2^2\bar D_{12} }
  \left(
  \frac{m_2^2}{\bar D_{2}}+\frac{m_{12}^2}{\bar{q}_{12}^2}
  \right) 
  -\left(
        \frac{1}{\bar D^3_1}
       -\frac{1}{\bar{q}_1^6}
    \right)
 \frac{q_1^2+2(q_1 \cdot q_2)}{\bar{q}_2^4 \bar{q}_{12}^2} 
\right\}, \nl
\eqa
where the denominator expansion in the r.h.s. is {\em the same one} performed in front of $q_1^\alpha q_1^\beta$ in Eq.~(\ref{eq:ex2l}). 
An explicit computation~\cite{Donati:2013voa} gives
\bqa
\int [d^4q_1] [d^4q_2] 
 \frac{\mu^2|_1}{\bar D_1^3\bar D_2\bar D_{12}} =
-\pi^4 \left(
\frac{1}{2}
+\frac{2}{3} f
\right),
\eqa
with
\bqa
\label{eq:f}
f = \frac{i }{\sqrt{3}}
\left(
 {\rm Li_2}\big(e^{ i \frac{\pi}{3}} \big)
-{\rm Li_2}\big(e^{-i \frac{\pi}{3}} \big)
\right).
\eqa
Integrals with powers of $\mu^2|_i$ in the numerators are called {\em extra integrals} and automatically generate the constants needed to preserve gauge invariance when decomposing tensors.\footnote{In Appendix \ref{appa} we demonstrate that FDR and DR tensors coincide at one loop and differ at higher orders.} For example\footnote{The global prescription {\em is not} applied to the r.h.s. of the first line in Eq.~(\ref{eq:tensdec}) because $q_1^2$ is generated by tensor decomposition and {\em not} by Feynman rules. Rather, the relation $q_1^2 = \qbar_1^2+\mu^2|_1$ is used to achieve the reduction given in the second line.}
\bqa
\label{eq:tensdec}
&&
\int [d^4q_1] [d^4q_2] 
  \frac{q^\alpha_1q^\beta_1}{\bar D_1^3\bar D_2\bar D_{12}} = \frac{g^{\alpha \beta}}{4}\!\!
\int [d^4q_1] [d^4q_2] 
 \frac{q_1^2}{\bar D_1^3\bar D_2\bar D_{12}} \nl
&&=
\frac{g^{\alpha \beta}}{4}\!\left(
\int [d^4q_1] [d^4q_2] 
 \frac{1}{\bar D_1^2\bar D_2\bar D_{12}} 
+
\int [d^4q_1] [d^4q_2] 
 \frac{m_1^2}{\bar D_1^3\bar D_2\bar D_{12}} 
+
\int [d^4q_1] [d^4q_2] 
 \frac{\mu^2|_1}{\bar D_1^3\bar D_2\bar D_{12}}\right). \nl
\eqa
An important consequence is 
\bqa
\label{eq:2looptens}
\int [d^4q_1] [d^4q_2] 
  \frac{4q^\alpha_1q^\beta_1-\qbar^2_1 g^{\alpha \beta}}{\bar D_1^3\bar D_2\bar D_{12}}
= - \pi^4 g^{\alpha\beta}
 \left(
\frac{1}{2}
+\frac{2}{3}f
\right) \ne 0.
\eqa

It is interesting to investigate how FDR integrals depend on $\mu$.\footnote{In the absence of IR divergences.}
The first term in the r.h.s. of Eq.~(\ref{eq:diff}) does not depend on $\mu$, because $\lim_{\mu \to 0}$ can be moved inside integration. On the other hand, any polynomially divergent integral in $J_{\rm INF}(q_1,\ldots, q_\ell)$ cannot contribute either, being proportional to positive powers of $\mu$, which vanish when $\mu \to 0$.  Therefore, the $\mu$ dependence of the l.h.s. is entirely due to powers of $\ln (\mu/\mur)$ generated by the logarithmically divergent subtracted integrals.
Therefore:
\begin{itemize}
\item[i)] FDR integrals depend on $\mu$ {\em logarithmically};
\item[ii)] if all powers of $\ln (\mu/\mur)$ are moved to the l.h.s. of Eq.~(\ref{eq:diff})\footnote{This is equivalent to a redefinition of FDR integration in which the powers of $\ln (\mu/\mur)$ {\em are not subtracted}. We assume this in the following.}, $\lim_{\mu \to 0}$ can be taken by formally trading $\ln (\mu)$ for
$\ln (\mur)$. 
\end{itemize}
Then, FDR integrals {\em do not depend on any cutoff} but only on $\mur$, which is interpreted as the renormalization scale.\footnote{See Sec.~\ref{sec:top}.}

In summary, higher-order calculations can be performed by interpreting the loop integrals as FDR ones. FDR directly produces renormalized Green's functions {\em with no need of an order-by-order renormalization}.\footnote{We emphasize that, at this stage, the QFT (although finite) is not predictive. It becomes so only when the free parameters in the Green's function are linked to experimental observables (see Sec.~\ref{sec:renandnonren}).} The reason is that algebraic manipulations in FDR integrals are allowed as if they where normal convergent loop integrals\footnote{As already observed, this preserves the symmetries of the QFT.} and, at the same time, no cutoff remains to be re-absorbed in the bare parameters of the Lagrangian, which is left untouched. This is in contrast with any  renormalization procedure based on counterterms, such as DR, where the presence of the cutoff at the intermediate stages of the calculation forces an iterative subtraction {\em \`a la} BPHZ~\cite{Bogoliubov:1957gp,Hepp:1966eg,Zimmermann:1969jj}. More in particular, FDR differs from the Zimmermann's definition of loop integration in three aspects:
\begin{itemize}
\item[i)] the FDR subtraction is obtained by a formal expansion of the original loop integrands around poles in ${\bar q^2_i}$, and not via a Taylor expansion in the external momenta;
\item[ii)] poles in ${\bar q^2_i}$ giving rise to UV divergences are subtracted without any attempt of re-introducing them into the Lagrangian;
\item[iii)]  gauge invariance is automatically respected via 
Eqs.~(\ref{eq:shift}) and (\ref{eq:canc}), while it must be enforced by hand in BPHZ.
\end{itemize}

Explicit examples of FDR calculations in renormalizable QFTs have been presented in~\cite{Pittau:2013qla,Donati:2013iya,Donati:2013voa}. Notably, the two-loop photon self-energy at the leading log and the two-loop ${\cal O}(\alpha_s)$ corrections to $H \to \gamma \gamma$ --\,mediated by an infinitely heavy top loop\,-- have been computed in~\cite{Donati:2013voa} without using order-by-order counterterms. It is precisely the absence of counterterms, together with the independence of the cutoff, which makes it appealing the use of FDR in non-renormalizable theories.\footnote{See Sec.~\ref{sec:noren}.}

\section{Classifying the UV divergences}
\label{sec:clas}
We devote this Section to a classification of the UV divergences contained in $J_{\rm INF}(q_1,\ldots, q_\ell)$.  By defining integrals as in Eq.~(\ref{eq:fdrdef}), and using the global prescription of Eq.~(\ref{eq:glopresc}), loop calculations can be carried out {\em without making any reference to $J_{\rm INF}(q_1,\ldots, q_\ell)$}.
Nevertheless, our classification provides an interesting physical interpretation of the discarded integrands. Thus, we adopt here the definition of FDR integration given in Eq.~(\ref{eq:diff}) and study the part subtracted from $J(q_1,\ldots, q_\ell)$.

We distinguish between scalars and irreducible tensors and show that the UV behavior of any QFT is completely parametrized in terms of a well defined set of logarithmically divergent scalar integrands depending only on $\mu$. 
We assume that 
polynomially divergences in $J_{\rm INF}(q_1,\ldots, q_\ell)$ do not contribute when $\mu \to 0$. This is trivially true if DR is used in the r.h.s. of Eq.~(\ref{eq:diff}), but it always holds\footnote{{\em Any regulator} can be used to evaluate Eq.~(\ref{eq:diff}).}, because polynomials in any cutoff scale drop in the difference.

\subsection{Scalars}
At the one-loop level just one logarithmically divergent integrand is responsible for all possible occurring UV infinities. Consider, in fact, the quadratically divergent integrand
\begin{equation}
\label{eq:i2}
\frac{1}{\dbar_{p_0}}
\end{equation}
where
\bqa
\label{defd}
\bar D_{p_i} = (q+p_i)^2-M_i^2 -\mu^2 \equiv \qbar^2-d_i~~~{\rm and}~~~p_0= 0,
\eqa
which also defines $d_i$.
It can be FDR expanded as follows
\bqa
\frac{1}{\dbar_{p_0}} &=& \left[\frac{1}{\bar q^2}\right] 
                + d_0 \left[\frac{1}{\bar q^4}\right]
                + \frac{d_0^2}{\bar D_0 \bar q^4}.
\eqa
The first term does not contribute, being quadratically divergent. Thus, the divergence is entirely generated by the integrand
\bqa
\label{eq:top1}
\left[\frac{1}{\bar q^4}\right],
\eqa
which is also responsible for the UV behavior of
\begin{equation}
\label{eq:i0}
\frac{1}{\bar D_{p_0}\bar D_{p_1}}.
\end{equation}
In fact
\bqa
\label{eq:i0a}
\frac{1}{\Dbar_{p_0} \Dbar_{p_1}} 
&=&
			\Bigg[\frac{1}{\qbar^4}\Bigg]
			+\frac{d_1}{\qbar^4\Dbar_{p_1}}
			+\frac{d_0}{\qbar^2\Dbar_{p_0}\Dbar_{p_1}}. 
\eqa

When the number of loops increases, the situation is similar. A two-loop example is given by the fundamental\footnote{All scalars of the type $1/({\bar D^\alpha_1\bar D^\beta_2\bar D^\gamma_{12}})$ are obtained by differentiating with respect to $m^2_1$, $m^2_2$ and $m^2_{12}$.} scalar integrand
\begin{equation}
J(q_1,q_2) = 
\frac{1}{\bar D_1\bar D_2\bar D_{12}}\,,
\end{equation}
with denominators written in Eq.~(\ref{eq:2lden}).
Its FDR expansion reads
\begin{eqnarray}
\label{eq:eq21}
J(q_1,q_2)  &=&
\left[\frac{1}{\bar{q}_1^2\bar{q}_2^2 \bar{q}_{12}^2}\right] \nonumber \\
&+& m_1^2
\left[
\frac{1}{\bar{q}_1^4\bar{q}_2^2 \bar{q}_{12}^2}
 \right]
+
 \frac{m_1^4}{(\bar D_1\bar{q}_1^4)}
\left[\frac{1}{\bar{q}_2^4} \right]
- m_1^4 \frac{q_1^2+2(q_1 \cdot q_2)}{(\bar D_1 \bar{q}_1^4)\bar{q}_2^4 \bar{q}_{12}^2}\nl
&+& m_2^2
\left[
\frac{1}{\bar{q}_1^2 \bar{q}_2^4\bar{q}_{12}^2}
 \right]
+
 \frac{m_2^4}{(\bar D_2\bar{q}_2^4)}
\left[\frac{1}{\bar{q}_1^4} \right]
- m_2^4 \frac{q_2^2+2(q_1 \cdot q_2)}{\bar{q}_1^4 (\bar D_2 \bar{q}_2^4) \bar{q}_{12}^2}\nl
&+&
 m_{12}^2
\left[
\frac{1}{\bar{q}_1^2 \bar{q}_2^2\bar{q}_{12}^4}
 \right]
+
 \frac{m_{12}^4}{(\bar D_{12}\bar{q}_{12}^4)}
\left[\frac{1}{\bar{q}_1^4} \right]
- m_{12}^4 \frac{q_{12}^2-2(q_1 \cdot q_{12})}{\bar{q}_1^4 \bar{q}_{2}^2
(\bar D_{12} \bar{q}_{12}^4)}
\nonumber \\
&+& 
 \frac{m_1^2 m_2^2}{(\bar D_1 \bar{q}_1^2)(\bar D_2 \bar{q}_2^2)\bar{q}_{12}^2} 
+\frac{m_1^2 m_{12}^2}{(\bar D_1 \bar{q}_1^2)\bar{q}_{2}^2(\bar D_{12} \bar{q}_{12}^2)} 
+\frac{m_2^2 m_{12}^2}{\bar{q}_{1}^2(\bar D_2 \bar{q}_2^2)(\bar D_{12} \bar{q}_{12}^2)} 
\nonumber \\
&+& 
\frac{m_1^2 m_2^2 m_{12}^2}{(\bar D_1 \bar{q}_1^2)(\bar D_2 \bar{q}_2^2)(\bar D_{12} \bar{q}_{12}^2)}\,, 
\end{eqnarray}
and since the first term in the r.h.s. is quadratically divergent, the only new logarithmic infinity is
\bqa
\label{eq:top2}
\left[
\frac{1}{\bar{q}_1^4\bar{q}_2^2 \bar{q}_{12}^2}
 \right].
\eqa 
At three loops, one obtains five additional logarithmic divergent 
integrands~\cite{Pittau:2012zd}
\bqa
\label{eq:top3}
&&
\left[
\frac{1}{\qbar_1^2\qbar_2^2\qbar_3^2 \qbar_{12}^2\qbar_{13}^2
((q_2-q_3)^2-\mu^2)} 
\right]\,,
\left[
\frac{1}{\qbar_1^2 \qbar_3^2 \qbar_2^4 \qbar_{12}^2\qbar_{23}^2} 
\right]\,,\nl
&&\left[
\frac{1}{\qbar_1^4 \qbar_2^2 \qbar_3^2 \qbar_{12}^2 \qbar_{123}^2} 
\right]\,,
\left[
\frac{1}{\qbar_1^4 \qbar_2^4 \qbar_3^2 \qbar_{123}^2} 
\right]\,,
\left[
\frac{1}{\qbar_1^6 \qbar_2^2 \qbar_3^2 \qbar_{123}^2} 
\right],
\eqa
and so on.
\subsection{Irreducible tensors}
\label{tensors}
The divergent integrands subtracted from tensors can be immediately read from their scalarized form.
For instance, at one-loop,
\bqa
\int[d^4q]\frac{q^\alpha q^\beta}{(\qbar^2-M^2)^3}= 
\frac{g^{\alpha \beta}}{4} \int[d^4q]\frac{1}{(\qbar^2-M^2)^2}.
\eqa
Thus, the UV part of the integrand reads
\bqa
\left\{\frac{q^\alpha q^\beta}{(\qbar^2-M^2)^3}\right\}_{\rm INF}= 
\frac{g^{\alpha \beta}}{4}  \Bigg[\frac{1}{\qbar^4}\Bigg],
\eqa
where the scalar in Eq.~(\ref{eq:top1}) appears again.

A two loop example is provided by Eq.~(\ref{eq:tensdec}). The first two terms of the second line contain the divergent scalar integrands dropped from its l.h.s.\footnote{They are obtained by differentiating Eq.~(\ref{eq:eq21}) with respect to $m_1^2$.}.
Therefore, as promised, the UV part of tensors is reducible to combinations of the same  divergent scalar integrands classified in the previous section.

\subsection{Topological renormalization}
\label{sec:top}
\begin{figure}
\begin{center}
\begin{picture}(300,180)(0,0)
\SetOffset(140,160)
\BCirc(-30,12){20}
\CCirc(-50,12){2}{}{}

\SetOffset(220,160)
\BCirc(-30,12){20}
\CCirc(-50,12){2}{}{}
\Line(-30,32)(-30,-8)

\SetOffset(100,90)
\BCirc(-30,12){20}
\Line(-30,32)(-30,12)
\Line(-30,12)(-15,-1)
\Line(-30,12)(-45,-1)

\SetOffset(180,90)
\BCirc(-30,12){20}
\Line(-40,30)(-40,-6)
\Line(-20,30)(-20,-6)

\SetOffset(260,90)
\BCirc(-30,12){20}
\CCirc(-50,12){2}{}{}
\Line(-30,18)(-30,-8)
\BCirc(-30,25){7}

\SetOffset(140,20)
\BCirc(-30,12){20}
\CCirc(-50,12){2}{}{}
\GOval(-30,12)(10,19)(90){1}
\CCirc(-40,12){2}{}{}

\SetOffset(220,20)
\BCirc(-30,12){20}
\GOval(-30,12)(10,19)(90){1}
\CCirc(-49,18){2}{}{}
\CCirc(-49,6){2}{}{}

\end{picture}
\caption{\label{fig:fig1} One-, two- and three-loop scalar topologies of the logarithmically divergent vacuum integrands. Dots denote denominators squared and $\mu\to 0$ is the only scale in the propagators.}
\end{center}
\end{figure}
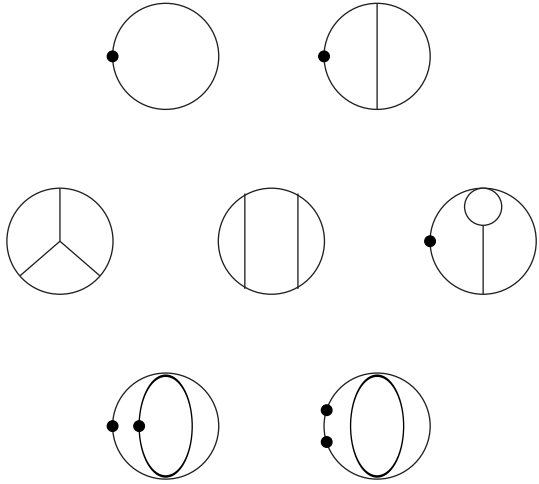

The classification of infinities given above can be rephrased in terms of vacuum topologies. The one-, two- and three-loop logarithmically divergent scalar integrands in Eqs.~(\ref{eq:top1}),~(\ref{eq:top2}) and~(\ref{eq:top3}) can be thought as the vacuum bubbles of Fig.~\ref{fig:fig1}.
An interpretation which legitimates their subtraction from the
original integrands is that they represent unphysical vacuum configurations, which do not take part in the scattering process~\cite{Pittau:2012zd}. Differently stated, the subtraction of $J_{\rm INF}$ in Eq.~(\ref{eq:diff}) can be understood as an order-by-order redefinition of the perturbative vacuum, which we dub {\em topological renormalization}, and FDR is an operative way to choose, order by order in the loop expansion, the {\em right} vacuum. 
Thus, the problem with the UV divergences is not where they can be accommodated, but rather what is the price of subtracting them.
As pointed out in Sec.~\ref{sec:fdr}, polynomially divergent integrands can be dropped {\em at no price}\footnote{This is why they are omitted in Fig.~\ref{fig:fig1}.}, since they naturally decouple, without any observable effect, in the limit $\mu \to 0$. Conversely, logarithmic divergent integrands cannot be fully subtracted, because, by doing so, logarithmic IR\footnote{The topologies in Fig.~\ref{fig:fig1} produce IR divergent singularities when $\mu\to 0$. Discarding such IR infinities has nothing to do with the UV subtraction.}  divergences would be generated (when $\mu \to 0$) in $J_{\rm F,\ell}$, which one interprets as the physical part of the interaction. Therefore, in order to keep $J_{\rm F,\ell}$ finite and independent on $\mu$, one introduces an (arbitrary) separation scale $\mur$ between the UV and the IR part and does not subtract the latter. This can be clearly exemplified in the context of the one-loop example of Eq.~(\ref{eq:i0a}), regularized with a momentum cutoff $\Lambda_{\rm UV}$.
The integral over the divergent contribution reads
\bqa
\label{eq:vac1loop}
% \lim_{\mu \to 0}\, 
\int_{\Lambda_{\rm UV}} d^4q \left[\frac{1}{\qbar^4}\right] = 
% \lim_{\mu \to 0}\,
%
-i \pi^2 \left(1+\ln \mu^2/\Lambda_{\rm UV}^2\right)= 
% \lim_{\mu \to 0}\,
%
-i \pi^2 \left(1+\ln \mu^2/\mur^2 +\ln  \mur^2/\Lambda_{\rm UV}^2 \right)
% 
%2 i\pi^2 \left(\int_0^{\mur} dq + \int_{\mur}^\Lambda dq \right)  \frac{q^3}{(q^2+\mu^2)^2}, 
\eqa
and, if $\ln \mu^2/\mur^2$ is not subtracted, the  dependence on $\mu$ of $J_{\rm F,\ell}$ is replaced by a dependence on $\mur$, which plays the role of the renormalization scale, so that $\lim \mu \to 0$ can be taken.
The DR version of Eq.~(\ref{eq:vac1loop}) is given
in Eq.~(\ref{eq:onlydiv1}), where the same logarithm 
appears.\footnote{Now the arbitrary scale of DR becomes the renormalization scale.}  Notice that, owing to the IR origin of $\ln \mu^2$, {\em any} UV regulator produces the same coefficient in front of the logarithm. Different constants generated by different UV regulators are immaterial because they are fully subtracted. This is again a consequence of the independence of FDR from any regularization procedure.

With all of that in mind, to renormalize a QFT at $\ell$-loops one simply drops the divergent integrands, computes the physical part $J_{\rm F,\ell}$ in the limit $\mu \to 0$ (to get its logarithmic dependence) and evaluate the result in $\mu= \mur$, which corresponds to the definition of FDR $\ell$-loop integral.

\section{Fixing renormalizable and non-renormalizable theories}
\label{sec:renandnonren}
In what follows, we consider the problem of fixing a QFT in the context of the
topological renormalization approach driven by FDR. Consider a theory described by a Lagrangian
\bqa
{\cal L}(p_1,\ldots,p_m)
\eqa
dependent on $m$ bare parameters $p_i$, with $i= 1:m$.
Before making any prediction, the $p_i$ must be fixed in terms of
$m$ measurements
\bqa
\label{eq:meas}
{\cal O}_{i}^{\rm EXP} = {\cal O}_{i}^{\rm TH,\,\ell-loop}(p_1,\ldots,p_m),
\eqa 
which determine them in terms of measured observables ${\cal O}^{\rm EXP}_{i}$ and corrections computed at the loop level $\ell$ one is working 
\bqa
\label{eq:fixing}
p^{\rm \ell-loop}_i({\cal O}^{\rm EXP}_{1},\ldots,  {\cal O}^{\rm EXP}_{m}) \equiv \bar p_i. \eqa
Both the $p_i$ and the $\bar p_i$ are {\em finite} in FDR.
Since the UV divergences have already been dropped, the issue of fixing the QFT is separated from the UV subtraction. Thus, a {\em global} determination of the bare parameters is possible: the $p_i$ do not need to be calculated iteratively in the perturbative expansion when inverting Eqs.~(\ref{eq:meas}).\footnote{This is due to the absence of counterterms.}  We refer to the procedure leading to Eqs.~(\ref{eq:fixing}) as a {\em global finite} renormalization.

Once the $\bar p_i$ are known, the QFT becomes predictive.
In the following two Sections, we discuss renormalizable and the non-renormalizable theories in turn.

\subsection{Renormalizable QFTs}
\label{sec:ren}
When ${\cal L}$ is renormalizable, the calculation of an independent 
observable
\bqa
{\cal O}_{m+1}^{\rm TH,\,\ell-loop}(\bar p_1,\ldots,\bar p_m)
\eqa
is a prediction of the theory --\,accurate at the $\ell^{th}$ perturbative 
order\,-- in which
\bqa
\label{eq:incut}
\frac{\partial\, {\cal O}_{m+1}^{\rm TH,\,\ell-loop}(\bar p_1,\ldots,\bar p_m)}{\partial \mur} = 0.
\eqa
As in any other renormalization scheme, independence of the renormalization scale is guaranteed up to $\ell$ loops because the {\em same} combinations of interactions with the {\em same} dependence on $\mur$ appear both in ${\cal O}_{m+1}^{\rm TH,\,\ell-loop}$ and in the r.h.s. of Eqs.~(\ref{eq:meas}). Thus, $\mur$ gets compensated when ${\cal O}_{m+1}^{\rm TH,\,\ell-loop}$ is calculated in terms of the $\bar p_i$. 

\subsection{Non-renormalizable QFTs}
\label{sec:noren}
If ${\cal L}$ is non-renormalizable, the global finite renormalization does not necessarily\footnote{
The absence of UV infinities in $J_{\rm INF}$ is a sufficient but not necessary condition for the absence of $\ln{\mur}$ in $\Black J_{{\rm F},\ell}$. 
Consider, for instance, the combination
\bqa
J(q_1,q_2)= \left(
 \frac{2}{\dbar^2_1\dbar_2\dbar_{12}}
-\frac{1}{\dbar^2_1 \dbar^2_2}
+\frac{4m^2}{\dbar^3_1\dbar^2_2}
\right),
\eqa
with $m^2_1= m^2_2= m^2_{12}= m^2$.
One computes
\bqa
\int [d^4q_1][d^4q_2]
J(q_1,q_2) = 2 \pi^4 f,
\eqa
with $f$ given in Eq.~(\ref{eq:f}). While
\bqa
\mur^{-2 \epsilon} \int d^nq_1 d^nq_2
J(q_1,q_2)_{\rm INF} 
= \pi^4
\left[
-2 \left({\frac{1}{\epsilon}}+\ln\pi + \gamma_E +\ln \frac{m^2}{\mur^2}
\right)-1+2 f
\right].
\eqa
} compensate the dependence on $\mur$ of ${\cal O}_{m+1}^{\rm TH,\,\ell-loop}$, which might depend on $\ln\mur$ at the {\em same} perturbative order one is computing
\bqa
\label{mp1}
{\cal O}_{m+1}^{\rm TH,\,\ell-loop}(\bar p_1,\ldots,\bar p_m,\ln{\mur}). 
\eqa
Since $\mur$ is not calculable in the framework of the theory itself, the QFT has to be considered as an effective one. However, it is possible to restore its predictivity in the infinite loop limit by determining $\mur$ from data.\footnote{Here the FDR replacement of $\mu \to \mur$ discussed in Sec.~\ref{sec:top} plays an essential role. The scale $\mur$ can be adjusted, while $\mu \to 0$.} 
In fact, combinations of observables in which $\mur$ disappears can still be unambiguously predicted. For instance, one computes, at one loop,
\bqa
{\cal O}_{m+1}^{\rm TH,\,1-loop}(\bar p_1,\ldots,\bar p_m,\ln{\mur}) &=& c_1 \ln{\mur} + d_1 \nl
{\cal O}_{m+2}^{\rm TH,\,1-loop}(\bar p_1,\ldots,\bar p_m,\ln{\mur}) &=& c_2 \ln{\mur} + d_2,
\eqa
and a quantity independent of $\mur$ can be constructed as follows
\bqa
{\cal O}_{\rm Predictable}^{\rm EXP}(\bar p_1,\ldots,\bar p_m) &=& 
 \frac{{\cal O}_{m+1}^{\rm EXP}(\bar p_1,\ldots,\bar p_m,\ln{\mur})}{c_1}-
\frac{{\cal O}_{m+2}^{\rm EXP}(\bar p_1,\ldots,\bar p_m,\ln{\mur})}{c_2} \nl
&=&\frac{d_1}{c_1}- \frac{d_2}{c_2},
\eqa
which is equivalent to extracting $\ln\mur$ from 
${\cal O}_{m+2}^{\rm EXP}= {\cal O}_{m+2}^{\rm TH,\,1-loop}$ and inserting the result in ${\cal O}_{m+1}^{\rm TH,\,1-loop}$. This can be generalized to $\ell$-loops. {\em One} additional measurement fixes $\mur$ {\em at any order}
\bqa
\label{eq:mp2}
{\cal O}_{m+2}^{\rm EXP}= {\cal O}_{m+2}^{\rm TH,\,\ell-loop}(\bar p_1,\ldots,\bar p_m, \ln \mur^\prime),
\eqa
and ${\cal O}_{m+1}^{\rm TH,\,\ell-loop}(\bar p_1,\ldots,\bar p_m,\ln{\mur^\prime})$ is a prediction of the non-renormalizable QFT in terms of the $m$ input observables ${\cal O}_i^{\rm EXP}$ of Eq.~(\ref{eq:meas}) and ${\cal O}_{m+2}^{\rm EXP}$.\footnote{Requiring perturbativity implies the condition
\bqa
|g^2\ln \mu^\prime_R| < 1
\eqa
on the solution $\mur^\prime$, where $g$ is the coupling constant of the QFT.
} 
The extra measurement in Eq.~(\ref{eq:mp2}) separates any UV effect from the physical degrees of freedom.\footnote{A similar procedure has been studied by Bettinelli, Ferrari and Quadri~\cite{Ferrari:2007zz} in the context of the nonlinear sigma model~\cite{Ferrari:2005ii} and of the massive Yang-Mills theory~\cite{Bettinelli:2007tq}. They obtained a predictive version of such non-renormalizable QFTs in DR by throwing away poles in  ${\epsilon}$ and fixing the renormalization scale with an additional measurement.}

It is important to realize that, within the proposed formulation, nothing can be said on the form of the Lagrangian. If ${\cal L}$  {\em does} contain an infinite amount of interactions, they have to be fixed {\em anyway}, either by knowing the underlying theory (as usually done when matching HQET and NRQCD with QCD~\cite{Manohar:1997qy,Brambilla:2004jw}) or by experiment. Nevertheless, at fixed number of terms in ${\cal L}$, all QFTs, including the non-renormalizable ones, can be tested experimentally at any loop order.
Unlike in the customary effective approach to non-renormalizable QFTs, there is no need, in FDR, to introduce new interactions to absorb UV infinities, and interactions other than those originally present in the bare Lagrangian are generated only by loops.

It is possible that QFTs exist for which an internal determination of their UV completion is intrinsically impossible or not at reach. The suggested procedure can then be used to rescue such theories in their minimal form. 
In other words, parametrizing unknown or uncomputable UV effects in terms of a 
{\em single} adjustable scale $\mur$ might be considered as a powerful and economic way to restore predictivity.
For instance, the minimal version of Einstein's gravity~\cite{Donoghue:1994dn}, if interpreted {\em \`a la FDR}, could already provide, at least effectively, a consistent and computable description of gravitation at the quantum level.
  
\section{Conclusions}
Based on the FDR classification of the UV infinities in terms of their topology, an alternative interpretation of the renormalization procedure can be formulated. 
Topological renormalization is introduced as an operative way to subtract the divergences directly at the level of the integrand of the loop functions, that, in turn, can be thought as an order-by-order redefinition of the perturbative vacuum.

This formulation is equivalent to the standard renormalization procedure for renormalizable theories -although technically simpler- and, in the case of non-renormalizable theories, it allows one to fit the scale at which the UV subtraction is performed. The net effect is that the bare Lagrangian is left untouched and
one additional measurement fixes the theory, which becomes predictive in the infinite loop limit. 
In this context, the validity of a Quantum Field Theory should by judged a-posteriori by comparing predictions with data and not necessarily based on its renormalizability.

\acknowledgments
We acknowledge discussions with Mohab Abou Zeid, David Kosower, Stefan Weinzierl, Einan Gardi, Simon Badger, Claude Duhr, Thomas Hahn, Pierpaolo Mastrolia, Francesco Tramonano, Edoardo Mirabella, Nora Brambilla and Antonio Vairo.
This work was supported by the European Commission through contracts ERC-2011-AdG No 291377 (LHCtheory) and PITN-GA-2012-316704 (HIGGSTOOLS). 
We also thank the support of the MICINN project FPA2011-22398 (LHC@NLO) and the Junta de Andalucia project P10-FQM-6552.

\appendix
\section{FDR versus Dimensional Regularization}
\label{appa}
One-loop DR tensors in $\overline{\rm MS}$ and FDR tensors coincide.
Consider, in fact, the integrand of a rank-two three-point function
\bqa
\label{eq:tens2}
\frac{q^\alpha q^\beta}{D_{p_0} D_{p_1} D_{p_2}}&=&
\lim_{\mu \to 0}
\frac{q^\alpha q^\beta}{\bar D_{p_0}\bar D_{p_1} \bar D_{p_2}}\nl
&=&
\lim_{\mu \to 0} \left\{
\frac{q^\alpha q^\beta}{\qbar^6}
+ q^\alpha q^\beta 
\left(
 \frac{d_2}{\qbar^6 \bar D_{p_2}}
+\frac{d_1}{\qbar^4 \bar D_{p_1} \bar D_{p_2}}
+\frac{d_0}{\qbar^2 \bar D_{p_0} \bar D_{p_1} \bar D_{p_2}}
\right)\right\}.
\eqa
 Upon DR integration 
\bqa
\frac{q^\alpha q^\beta}{\qbar^6}
\eqa
can be replaced by
\bqa
\frac{g_{\alpha\beta}}{4} \frac{1}{\qbar^4}.
\eqa
This property is a consequence of the one-loop gauge preserving conditions of DR~\cite{Wu:2003dd}, and it holds for any logarithmically divergent one-loop integral, e.g.
\bqa
 \int d^nq\, \frac{q^\alpha q^\beta q^\rho q^\sigma }{\qbar^8} 
= \frac{(g^{\alpha \beta}g^{\rho\sigma}+g^{\alpha \rho}g^{\beta \sigma }+g^{\alpha \sigma}g^{\beta \rho})}{24} \int d^nq\, \frac{1}{\qbar^4}.
\eqa 
Thus, all UV infinities can be made proportional to the DR integral
\bqa
\label{eq:onlydiv1}
\mur^{-\epsilon }\int d^nq\,\frac{1}{\qbar^4}
= i \pi^2 \left(
-\frac{2}{\epsilon}-\gamma_E-\ln \pi -\ln \frac{\mu^2}{\mur^2}
\right),
\eqa
and since all terms of the previous Equation but $\ln \frac{\mu^2}{\mur^2}$ are subtracted both in DR (in $\overline{\rm MS}$) and FDR, DR and FDR one-loop tensors represent the same mathematical object. For instance
\bqa
\mur^{-\epsilon }\int d^nq\, 
\frac{4q^\alpha q^\beta -q^2 g^{\alpha \beta}}{D_{p_0} D_{p_1} D_{p_2}}
=  \frac{i \pi^2}{2} g^{\alpha \beta} =
\int [d^4q]\, 
\frac{4q^\alpha q^\beta -\qbar^2 g^{\alpha \beta}}{\bar D_{p_0}\bar D_{p_1} \bar D_{p_2}}.
\eqa

A corollary of this theorem is the equivalence, at one loop, between FDR and Dimensional Reduction\footnote{Where no explicit $\epsilon$ appears in the numerator of the loop integrals.} in the  $\overline{\rm MS}$ scheme.

At $\ell$ loops, with $\ell > 1$, DR tensors differ from their FDR counterpart. For example
\bqa
\mur^{-2 \epsilon } \int d^nq_1 d^nq_2 
  \frac{4q^\alpha_1q^\beta_1-q^2_1 g^{\alpha \beta}}{D_1^3 D_2 D_{12}}
= \frac{\pi^4}{2} 
\left( 
\frac{1}{\epsilon}+\gamma_E+\ln \pi
+\ln \frac{m_1^2}{\mur^2} -\frac{1}{4} 
\right) g^{\alpha\beta}, 
\eqa
which does not match Eq.~(\ref{eq:2looptens}).
This is due to the presence of the cutoff $\epsilon$. In DR physical results are obtained only after poles and constants created by tensor decomposition are consistently combined with lowest order counterterms required by the order-by-order renormalization.\footnote{DR and FDR coincide at one loop owing to the lack of lowest order counterterms.}  
Conversely, FDR tensors are regulator free objects and, thanks to the absence of counterterms, physical constants are solely generated by extra integrals via tensor decomposition.
 
\bibliography{onthe}{}

\providecommand{\href}[2]{#2}\begingroup\raggedright\begin{thebibliography}{10}

\bibitem{Bogoliubov:1957gp}
N.~Bogoliubov and O.~a. Parasiuk, {\it {On the Multiplication of the causal
  function in the quantum theory of fields}},  {\em Acta Math.} {\bf 97} (1957)
  227--266.

\bibitem{Hepp:1966eg}
K.~Hepp, {\it {Proof of the Bogolyubov-Parasiuk theorem on renormalization}},
  {\em Commun.Math.Phys.} {\bf 2} (1966) 301--326.

\bibitem{Zimmermann:1969jj}
W.~Zimmermann, {\it {Convergence of Bogolyubov's method of renormalization in
  momentum space}},  {\em Commun.Math.Phys.} {\bf 15} (1969) 208--234.

\bibitem{Dyson:1949ha}
F.~Dyson, {\it {The S matrix in quantum electrodynamics}},  {\em Phys.Rev.}
  {\bf 75} (1949) 1736--1755.

\bibitem{Dyson:1949bp}
F.~Dyson, {\it {The Radiation theories of Tomonaga, Schwinger, and Feynman}},
  {\em Phys.Rev.} {\bf 75} (1949) 486--502.

\bibitem{Salam:1951sm}
A.~Salam, {\it {Overlapping divergences and the S matrix}},  {\em Phys.Rev.}
  {\bf 82} (1951) 217--227.

\bibitem{Salam:1951sj}
A.~Salam, {\it {Divergent integrals in renormalizable field theories}},  {\em
  Phys.Rev.} {\bf 84} (1951) 426--431.

\bibitem{Weinberg:1978kz}
S.~Weinberg, {\it {Phenomenological Lagrangians}},  {\em Physica} {\bf A96}
  (1979) 327.

\bibitem{Pittau:2012zd}
R.~Pittau, {\it {A four-dimensional approach to quantum field theories}},  {\em
  JHEP} {\bf 1211} (2012) 151, [\href{http://xxx.lanl.gov/abs/1208.5457}{{\tt
  arXiv:1208.5457}}].

\bibitem{Pich:1998xt}
A.~Pich, {\it {Effective field theory: Course}},
  \href{http://xxx.lanl.gov/abs/hep-ph/9806303}{{\tt hep-ph/9806303}}.

\bibitem{Pittau:2013qla}
R.~Pittau, {\it {QCD corrections to $H \to gg$ in FDR}},  {\em Eur.Phys.J.}
  {\bf C74} (2014) 2686, [\href{http://xxx.lanl.gov/abs/1307.0705}{{\tt
  arXiv:1307.0705}}].

\bibitem{Donati:2013iya}
A.~M. Donati and R.~Pittau, {\it {Gauge invariance at work in FDR: $H \to
  \gamma \gamma$}},  {\em JHEP} {\bf 1304} (2013) 167,
  [\href{http://xxx.lanl.gov/abs/1302.5668}{{\tt arXiv:1302.5668}}].

\bibitem{Donati:2013voa}
A.~M. Donati and R.~Pittau, {\it {FDR, an easier way to NNLO calculations: a
  two-loop case study}},  {\em Eur.Phys.J.} {\bf C74} (2014) 2864,
  [\href{http://xxx.lanl.gov/abs/1311.3551}{{\tt arXiv:1311.3551}}].

\bibitem{Sterman:1994ce}
G.~F. Sterman, {\em An Introduction to quantum field theory}.
\newblock Cambridge University Press, 1994.

\bibitem{Collins:1984xc}
J.~C. Collins, {\em Renormalization}.
\newblock Cambridge University Press, 1984.

\bibitem{veltman74}
M.~Veltman, {\it {Gauge Field Theories}},  {\em Proceedings of the VI
  International Symposium on Electron and Proton Interaction at High Energies}
  (1973) 429--447.

\bibitem{Ferrari:2007zz}
R.~Ferrari, D.~Bettinelli, and A.~Quadri, {\it {A new approach to
  nonrenormalizable theories}},  {\em Proceedings of 21st Les Rencontres de
  Physique de la Vallee d'Aoste} (2007) 551--572.

\bibitem{Ferrari:2005ii}
R.~Ferrari, {\it {Endowing the nonlinear sigma model with a flat connection
  structure: A Way to renormalization}},  {\em JHEP} {\bf 0508} (2005) 048,
  [\href{http://xxx.lanl.gov/abs/hep-th/0504023}{{\tt hep-th/0504023}}].

\bibitem{Bettinelli:2007tq}
D.~Bettinelli, R.~Ferrari, and A.~Quadri, {\it {A Massive Yang-Mills Theory
  based on the Nonlinearly Realized Gauge Group}},  {\em Phys.Rev.} {\bf D77}
  (2008) 045021, [\href{http://xxx.lanl.gov/abs/0705.2339}{{\tt
  arXiv:0705.2339}}].

\bibitem{Manohar:1997qy}
A.~V. Manohar, {\it {The HQET / NRQCD Lagrangian to order alpha / m-3}},  {\em
  Phys.Rev.} {\bf D56} (1997) 230--237,
  [\href{http://xxx.lanl.gov/abs/hep-ph/9701294}{{\tt hep-ph/9701294}}].

\bibitem{Brambilla:2004jw}
N.~Brambilla, A.~Pineda, J.~Soto, and A.~Vairo, {\it {Effective field theories
  for heavy quarkonium}},  {\em Rev.Mod.Phys.} {\bf 77} (2005) 1423,
  [\href{http://xxx.lanl.gov/abs/hep-ph/0410047}{{\tt hep-ph/0410047}}].

\bibitem{Donoghue:1994dn}
J.~F. Donoghue, {\it {General relativity as an effective field theory: The
  leading quantum corrections}},  {\em Phys.Rev.} {\bf D50} (1994) 3874--3888,
  [\href{http://xxx.lanl.gov/abs/gr-qc/9405057}{{\tt gr-qc/9405057}}].

\bibitem{Wu:2003dd}
Y.-L. Wu, {\it {Symmetry preserving loop regularization and renormalization of
  QFTs}},  {\em Mod.Phys.Lett.} {\bf A19} (2004) 2191--2204,
  [\href{http://xxx.lanl.gov/abs/hep-th/0311082}{{\tt hep-th/0311082}}].

\end{thebibliography}\endgroup
\bibliographystyle{JHEP}
\end{document}